**Revisiting money and labor for valuing environmental goods and services in developing countries**

Habtamu Tilahun Kassahun[1,†], Jette Bredahl Jacobsen[2], Charles F. Nicholson[3,4]

[1]Australian Rivers Institute, Griffith University, Nathan, Australia

[2]Department of Food and Resource Economics and Centre for Macroecology, evolution and climate, University of Copenhagen, Denmark

[3]Charles H. Dyson School of Applied Economics and Management, Cornell University, Ithaca, USA

[4]Nijmegen School of Management, Radboud University, Nijmegen, Netherlands

[†] **Corresponding author**: h.kassahun@griffith.edu.au

- The published vision is available at http://dx.doi.org/10.1016/j.ecolecon.2020.106771

**Citation**

**Abstract**

Many Stated Preference studies conducted in developing countries provide estimates of a low willingness to pay (WTP) for a wide range of goods and services. However, recent studies in these countries indicate that this may partly be a result of the choice of payment vehicle, not the preference for the good. Thus, low WTP may not indicate a low welfare effect for public projects in developing countries. We argue that in a setting where 1) there is imperfect substitutability between money and other measures of wealth (e.g., labor), and 2) institutions are perceived to be corrupt, including payment vehicles that are currently available to the individuals and less prone to corruption may be needed to obtain valid welfare estimates. Otherwise, we risk underestimating the welfare benefit of projects. We demonstrate this through a rural household contingent valuation (CV) survey designed to elicit the value of access to reliable irrigation water in Ethiopia. Of the total average annual WTP for access to reliable irrigation service, cash contribution comprises only 24.41%. The implication is that socially desirable projects might be rejected based on cost-benefit analysis as a result of welfare gain underestimation due to mismatch of payment vehicles choice in valuation studies.

**1. Introduction**

The application of stated preference (SP) methods such as contingent valuation (CV) and choice experiments (CE) for valuing environmental goods and services is growing in developing countries in recent years (Alemu et al., 2017; Bennett and Birol, 2010; Kassahun et al., 2016; Kassahun et al., 2020b; Meginnis et al., 2020; Navrud and Vondolia, 2019; Rai and Scarborough, 2015; Rakotonarivo et al., 2017; Tilahun et al., 2017; Vondolia and Navrud, 2019; Zemo et al., 2019). The results of valuation studies provide valuable information for the decision to the provision of these goods and services. An issue that remains relatively unexplored is that many SP studies conducted in developing countries provide estimates of a low willingness to pay (WTP) for a wide range of goods and services in absolute and comparison to the cost of provision (Whittington, 2010). Low WTP could suggest that the good or service under valuation is not a priority for many of the respondents in developing country settings. However, recent SP studies in developing countries indicate the need for a more careful interpretation of low WTP estimates (Abramson et al., 2011; Kassahun, 2014; Kassahun et al., 2020a).

Low WTP may not indicate a low demand for public projects in developing countries. One issue concerns public trust in the development and implementation of environmental goods and services (Birol and Das, 2012; Kassahun et al., 2020a; Oh and Hong, 2012). Kassahun et al. (2020a) showed that the standard assumption that WTP for public goods increases with income failed in the presence of high instructional distrust. Similarly, Chen and Hua (2015) have reported about 62 % of the sampled respondents refused to pay for a project because of distrust in the government in China. Distrust is not only limited to a government; for managing common-pool resources, Kassahun et al. (2020b) demonstrate that farmers with high levels of doubt about cooperation among themselves have a lower WTP for integrated watershed management. As a result, farmers would require an alternative means of incentive to participate in that land management program. Thus, trust issues, often prevalent in developing-country contexts, should be carefully accounted for in the design of the study and also in the analysis before making inferences about WTP.

Another issue is the form in which payments for goods and services made. In order to obtain a valid welfare estimate through a stated preference study, choosing a widely used payment vehicle for trading in a market is required. Furthermore, the present bids, the coverage of the payment vehicle, and the organization that managing the funds must be in the acceptable levels (Hassan et al., 2018; Morrison et al., 2000). If cash is infrequently used, and a society transacts primarily through bartering, using money to value goods and services may not be appropriate. When the cash economy is of limited importance as in rural areas of developing countries, WTP estimates based on monetary contributions alone may result in understated welfare effects for environmental goods and services (Abramson et al., 2011; Gibson et al., 2016; Schiappacasse et al., 2013). As a result, many researchers have used alternative payment vehicle systems for welfare measurement in valuation studies, often labor contributions[1].

Nevertheless, labor contributions are not a perfect substitute for cash payments (Kassahun and Jacobsen, 2015; Vondolia and Navrud, 2019), for two main reasons. First, labor can lack sufficient coverage as a payment vehicle. Labor contributions may not be a feasible option when the location of environmental management activities are far away from the residential area of the respondent. Thus, the ones who benefit may not legally nor geographically be the ones who can provide labor. Second, because of the potential difference in the availability of cash or labor, some individuals may not want to/be able to contribute to labor, preferring cash contributions, and vice versa. Thus, including both payment vehicles in a valuation study may capture the demand for an environmental good more accurately by allowing the respondent more flexibility to reveal true preferences with their choice of payment vehicle. The use of multiple payment vehicles is particularly important due to the decline of cash usage as a dominant exchange instrument in rural regions developing countries far from major urban centers (Pankhurst, 2007). However, the literature on CV methods that combine labor and money to elicit value in developing countries has overlooked three potential issues.

The first issue is the sequencing questions about two payment vehicles in the CV. Regardless of the sequence, it is common to estimate independent WTP and willingness to contribute labor (WTC) models assuming that the distribution of the sequence of choice between two payment vehicles is exogenous. Numerous studies employ this approach, for example, Hung et al. (2007) for forest fire prevention, Kassahun (2009) for the reliability of irrigation service, Casiwan-Launio et al. (2011) for sustainable marine fisheries, Tilahun et al. (2015) for forest conservation, and Tilahun et al. (2017) for prevention of expansion of alien tree species. However, this approach can result in biased coefficients and welfare estimates due to the correlation of the unobservable components of utility across the sequence of valuation questions. This is a case similar to the issue of correlation in double-bounded CV method with the same (money) payment vehicle that Alberini et al. (1997) and Poe et al. (1997) emphasized in early applications of CV. Yet, it has been largely ignored in the CV literature that combines money and labor as a payment vehicle.

---

[1] Section 2 provides an overview of the valuation of environmental goods and services for the last 30 years using both money and labor in developing countries.

Second, in dichotomous choice CV, the trade-off between two payment vehicles for two consecutive valuation questions is conditional on the first valuation equation. CV differs from CE, which allows respondents to the trade-off between different payment vehicles for each repeated valuation question (Kassahun and Jacobsen, 2015; Kassahun et al., 2020b; Meginnis et al., 2020; Rai and Scarborough, 2013). Thus, capturing trade-offs across the sequence of choices between two payment vehicles requires the use of the response from the first valuation question (lagged payment vehicle) as a determinant of the second valuation question. Doing this captures not only trade-offs between payment vehicles but also avoids potential omitted variable bias. However, including a response from the first valuation question as a determinant of the second valuation question also raises a potential endogeneity problem that requires care.

The third issue is related to behavioral response anomalies (Johnston et al., 2017). One behavioral response anomaly is related to the inconsistency of responses with standard assumptions (Johnston et al., 2017). It could be a strategic or protest response. The valuation question order might influence strategic response in the case of different payment vehicles. For example, question order may pressure respondents to indicate a false or strategic value for the second payment vehicle following the rejection of the first payment vehicle, independent of the respondent's income and labor status. If this is the case, it distorts the value of the environmental service under consideration. Thus, further analyses of relative and absolute income and labor differences between respondents are required to identify whether the trade-off is determined by the question order effect or by the opportunity cost of labor or a combination of the two.

A second behavioral response anomaly is a similar concept with starting point bias in a bidding game or a repeated choice CV method with monetary payment vehicles (Boyle et al., 1985; Flachaire and Hollard, 2007; Herriges and Shogren, 1996). However, this is ignored in the literature when there are different payment vehicles. Starting point bias or anchoring is a biased estimate close to the starting value in a repeated choice. It could be a result of poorly defined valuation scenarios, or it could be a behavioral response that respondents do not want to reveal their real preference (Boyle et al., 1985; Johnston et al., 2017). For inference about WTP/WTC, a questionnaire pre-test to identify and minimize anchoring effect and post data collection investigation about the anchoring effect should be in place.

Given the preceding, we argue that in the absence of widely used payment vehicles and imperfect substitutability between money and other payment vehicles, including different payment vehicles, is needed to obtain valid welfare estimates. Using a single payment vehicle may lead to a biased welfare estimate because of exclusions of individuals who do not use that payment vehicle for trading purposes. Furthermore, we investigate the consequences of the payment vehicle exogeneity assumption and endogeneity on utility and parameter estimates. We demonstrate this through a rural household CV survey designed to elicit the value of access to reliable irrigation water in Ethiopia. Our findings highlight the importance of accounting for cross-payment-vehicle correlation and potential endogeneity biases that arise in the sequence of WTP and Willingness to contribute (WTC) valuation questions. The implication is also valid for a choice experiment (CE) survey that allows the respondent to choose payment vehicles and administer the CE survey using the selected payment vehicle. The remaining sections of the article are as follows: Section 2 describes the method; Section 3 presents the data; Section 4 describes and discusses the results. Finally, Section 5 provides brief concluding remarks.

## 2. Overview of money and labor for valuation of environmental goods and services

Starting from the work of Swallow and Woudyalew (1994), the assessment of labor contributions has received increasing attention in the valuation of environmental goods and services in developing countries. We reviewed literature from 1994 to 2020 that combined money with labor in various developing countries (Table 1) in a stated preference elicitation format. Application areas include land management, tree planting, fire prevention, weed control, animal disease control, irrigation service, flood risk reduction, water quality improvement, drought insurance, and maintaining fishery reserve. In these valuation studies, the addition of labor is inspired by three observations. First, a large share of rural communities does not supply labor to the market for wages or salaries (Barrett et al., 2008; Gibson et al., 2016)[2] . Thus, having cash may not necessarily grant access to labor for project implementation. An excellent example of this is a rural development project in Kenya, whose community members required to contribute labor are not allowed to substitute cash with labor even if they have cash (Echessah et al., 1997; Thomas, 1980). Second, in some cases, even monetary contribution is considered unrealistic for labor intensive projects, for example, see Hung et al. (2007) for fire prevention valuation project in Vietnam. Third, cash availability is also deficient in most rural, remote areas of developing countries (Gibson et al., 2016; Schiappacasse et al., 2013). Thus, labor contribution is advocated as an alternative means of valuing the demand for public projects for labor-intensive projects.

---

[2] Understanding of labor supply behavior in rural, remote regions of the world are essential for applied work. For example, in rural highlands of Ethiopia, where about 85% of the population dwell, three severe complications afflict empirical work on labor supply behavior. First, a large share of farmers does not supply labor to the market for wages or salaries. They engage in self-employment in subsistence agriculture. Second, a large part of the trading takes place as an exchange economy, locally known as the "Debo System" for labor contribution. The Debo system is an old social norm that individuals share their labor to help other farmers. In turn, the one who gets the help will return it when someone in his/her group needs it. This system does not involve cash. In this case, establishing reliable substitutability cash with labor is not possible, because the real market wage rate is unknown. Third, in the absence of market wage rate information, the welfare-maximizing labor allocation equilibrium conditions of allocative efficiency between the marginal revenue product of labor ( $MRP_L$ ) and market wage rate ( $w$ ) for the same labor, $MRP_L = w$, can be distorted ( Barrett et al. 2009).

**Table 1:** Scholarly articles that use labor or labor combined with money for valuation of different environmental goods and services in developing countries

| Authors | Methods | Area of application and Country |
|---|---|---|
| Abramson et al. (2011) | CE | Rural water service improvement, Zambia |
| Amare et al. (2016) | CV | Restoring indigenous tree species, Ethiopia |
| Asrat et al. (2004) | CV | Soil conservation, Ethiopia |
| Casiwan-Launio et al. (2011) | CV | Existence of fishery reserve, Philippines |
| Echessah et al. (1997) | CV | Tsetse Control, Kenya |
| Gibson et al. (2016) | CE | Improved drinking water quality, Cambodia. |
| Hung et al. (2007) | CV | Forest fire prevention, Vietnam |
| Kassahun and Jacobsen (2015) | CE | Integrated land management, Ethiopia |
| Kassahun et al. (2020b) | CE | Integrated land management, Ethiopia |
| Meginnis et al. (2020) | CE | Access to improved water service, Uganda |
| Navrud and Vondolia (2019) | CE | Reductions in flood risk, Ghana |
| Rai and Scarborough (2013) | CE | Mitigation of plant invaders, Nepal |
| Rai and Scarborough (2015) | CE | Mitigation of the invasive vine, Nepal |
| Schiappacasse et al. (2013) | CV | Forest restoration, Chile |
| Swallow and Woudyalew (1994) | CV | Animal disease control, Ethiopia |
| Tadesse et al. (2017) | CE | Drought insurance, Ethiopia |
| Tilahun et al. (2015) | CV | Plant Conservation, Ethiopia |
| Tilahun et al. (2017) | CV | Weed control, Ethiopia |
| Vondolia et al. (2014) | CV | Irrigation channel maintain, Ghana |
| Vondolia and Navrud (2019) | CE | Flood risk reduction, Ghana |

This literature indicates that both CV and CE are used in the studies. In recent years, CE has become more popular among practitioners due to its flexibility for allowing tradeoff between payment vehicles and other attributes within a single choice task (Kassahun et al., 2020b; Rai and Scarborough, 2013). However, adding two payment vehicles with other attributes of interest in a choice task may lead to a concern about choice complexity for surveyed individuals (Rai and Scarborough, 2015)[3]. Choice complexity is known for increasing the choice of status-quo alternative in developed countries for individuals with lower education status (Boxall et al., 2009). Concerned with choice complexity, Rai and Scarborough (2015) propose an alternative means of elicitation of values. In their CE study, if a respondent refused to contribute in monetary terms, the respondent is asked to contribute with labor. Their analysis did not address the sample selection issue raised by labor contribution only being asked to those refused to pay, an issue previously raised as necessary in the CV literature.

A vital consideration of designing a stated preference study is the importance of familiarity with payment vehicles and respondents' perception of the payment vehicle itself. Recently, Vondolia and Navrud (2019) used a split-sample design to test the effect of alternative payment vehicles on the purchase of environmental goods and services. Their result shows that the responses from non-monetary payment vehicles point out high uncertainty (low scale parameter estimates) compared to monetary payment vehicles. The authors suggest high uncertainty might be associated with less familiarity with non-monetary payment vehicles than monetary payment vehicles. Furthermore, high uncertainty might be linked to price volatility for payment vehicles with agricultural commodities, or it could be the difference in the opportunity cost of labor. To reduce the uncertainty associated with payment vehicles, Vondolia and Navrud (2019) recommended precision in the framing of valuation scenarios for non-monetary payment vehicles. Precision in the framing of valuation scenarios can be achieved through careful pre-test and focus group discussions and following some of the recent SP guidelines (Johnston et al., 2017).

---

[3] Choice complexity is not only about the payment vehicles but also the number of other attributes. If there are few attributes, choice complexity may not be a problem. Perhaps research might be needed to determine the optimal number of attribute labels safely administer in CE in a rural context.

In the following section, we present the case study and describe how the payment vehicles are designed in the valuation study to provide a context for the method, which we present in sequence.

3. **Case study**

We obtained data for this study from irrigation beneficiary farmers in Koga Watershed of the Upper Blue Nile Basin part of Ethiopia. The watershed is located South of Lake Tana in an area with high erosion rates (Gebrehiwot et al., 2010; Reynolds, 2012). The newly constructed irrigation reservoir can irrigate 7,000 hectares of land and extends to seven administrative districts (Kebele). However, during the survey period, irrigation had not yet begun. Nevertheless, irrigation farming is not new to the region—some farmers who have land adjacent to the Koga River practice irrigation farming (Kassahun et al., 2016). Kassahun et al. (2016) use part of the data to show how irrigation users' expectations about future productivity affect WTP considering a substantial proportion of individuals without prior irrigation farming experience. Labor contribution was not part of their analysis.

The data are from 210 randomly selected irrigation beneficiary household heads that interviewed from July to October 2008. Consistent with other locations in Ethiopia, more than 90% of household heads surveyed were male[4]. The sample represents approximately 12 percent of the households from two randomly selected districts. The primary purpose of the survey was to value the reliability of irrigation services both in terms of monetary and labor contributions.

---

[4] Thus, sample structure limits the ability to analyze WTP and WTC disaggregated by gender.

## 3.1. CV data

The CV study involves directly asking farmers how much they would be willing to pay or work in exchange for access to reliable irrigation services. In the valuation scenario, the potential threat of reservoir and irrigation channel sedimentation, and the need for appropriate soil conservation measures both in the upstream and downstream parts of the watershed informed to farmers. The valuation scenario, as translated from Amharic, is presented in the Text Box 1.

---

**Text box 1: Valuation scenario**

*Access to year-round irrigation water supply requires maintaining the health of the dam and irrigation channels from sedimentation in Koga watershed. This activity requires substantial work both in the upstream and downstream regions of the Koga watershed. In the upstream part, to reduce the amount of siltation to the downstream reservoir and irrigation channel, appropriate soil and water conservation work should be in place. However, because of its distance from the irrigation users' residential area, irrigation users cannot accomplish conservation activities and follow-up in the upper catchment. Therefore, local communities in the upper watershed should work on conservation activities. However, no one can force upstream residents to practice soil and water conservation activities for the benefit downstream of irrigation users. Therefore, to encourage the participation of upstream households in soil and water conservation activities, financial incentives are required. In the areas near irrigation users' residences or the downstream part of Koga watershed, it is possible to manage and maintains the dam and common irrigation channels by irrigation beneficiaries. Therefore, to access long-term, year-round irrigation water, irrigation beneficiary households often contribute money and labor time to maintain the health of the dam and irrigation channels.*

---

Then, following adjusting concerns that we detect during our focus group discussion and pre-test, we introduced the valuation questions in a sequence. During the focus group discussion and pre-test, we found that farmers have a distrust of a government institution (Kassahun, 2014). Farmers suggested the money to be collected and managed through their own newly established irrigation users' cooperative rather than the government institutes. Accordingly, we adjusted the valuation questions to reduce a protest response and increase consequentiality. The wording of the valuation questions is as follows as presented to the irrigation beneficiary farmers (Text Box 2).

---

**Text box 2: Valuation questions**

I.  *If you are provided access to irrigation water, will you vote for irrigation cooperative rules and regulation that will create a fund, if its passage will require all irrigation users to contribute (\_\_\_) ETB/household/Year/ kada[1] of land to keep the health of the dam and common irrigation channels to get year round reliable irrigation water supply? (Yes, No)*

II. *What is the maximum amount that you are willing to pay per kada of land for such a project per year for ten years? \_\_\_\_*

III. *In addition to cash contribution, if you are requested to contribute (\_\_\_\_) labor days per Kada of land per month to maintaining the health of the dam and irrigation channels from sedimentation to get year round irrigation water supply, are you willing to contribute, if its passage require all irrigation users to contribute? (Yes, No)*

IV. *What is the maximum number of days that you are willing to contribute for such a project per month for ten years? \_\_\_*

---

In the dichotomous choice (DC) valuation questions (I and III), the bid values for money[5] and labor[6] were randomly assigned to each respondent. We tested the range of the bids through focus group discussions and pre-test. The purpose of the open-ended follow-up valuation questions (II and IV) was to diagnose different inconsistent responses together with the DC response valuation question. First, we considered responses inconsistent if the respondent states a lower amount of labor or money in the follow-up valuation questions compared to DC valuation questions. In this case, we remove responses from the analysis. Accordingly, we removed 16 responses from the analysis. Second, we use the follow-up valuation questions to diagnose behavioral anomaly responses such as starting point bias/anchoring effect and strategic or protest responses. We test significant differences between means of maximum WTP/WTC from the follow-up question grouped by starting bids from the dichotomous CV questions to diagnoses starting point bias within the same payment vehicle. To test cross-payment vehicle starting point bias, we categorize the maximum willingness to work values by the bid price. Moreover, we check income and labor differences grouped by the responses of dichotomous CV questions to diagnose the presence of a strategic or protest response.

---

[5] Randomly assigned bid prices (*BidCash*): 25, 31, 37, 43, 49, 58 and 70 ETB/year/Kada of irrigable land.
[6] the randomly bid working days (*BidLabor*): 1, 1.5, 2, 2.5, and 3 days/month/Kada of irrigable land.

Furthermore, for individuals who refused to pay a monetary contribution or a labor contribution, we asked a separate question to identify the reasons for refusal[7]. We have also collected data on socioeconomic and demographic variables hypothesized to have an impact on WTP/WTC for reliable irrigation service. We will present the result following the method section.

## 4. Method

Correlation between unobservable and observable components of utility is a well-documented issue in discrete choice analysis with a repeated response. Accounting for this effect is a routine procedure in CE and double-bounded dichotomous choice CV for the same

---

[7] The question is worded as follows: If you are not willing to pay any amount, please identify your reason/s:
1. I cannot afford to pay.
2. I think the government should finance watershed management activities.
3. I do not believe conservation and management activities will result in a more reliable water supply.
4. I do not fully understand the question.
5. Other reasons, please identify _______.

If you are not willing to contribute any labor, please tell me your reason/s?

payment vehicle (Alberini et al., 1997; Poe et al., 1997; Train, 2009). Similar methodological approaches can be applied to test the exogeneity of a sequence of two different dichotomous choice CV questions using a bivariate probit model. The bivariate probit model addresses potential endogeneity issues without the need for instrumental variables for one endogenous variable (Filippini et al., 2018; Heckman, 1978; Maddala, 1983; Martínez-Espiñeira and Lyssenko, 2011; Monfardini and Radice, 2008).

If we assume that the utility of an individual, i, for reliable irrigation service valued in a sequence of valuation questions with money, $U_{1i}$, and labor, $U_{2i}$, then 1 is the valuation question in cash contributions and 2 in labor contributions. Thus, the utility for reliable irrigation service given by:

$$U_{1i} = v_{1i} + \varepsilon_{1i} = \alpha + \beta_1 BidCash_{1i} + \sum_{\beta=2}^{B} \beta x_i + \varepsilon_{1i}$$
$$U_{2i} = v_{2i} + \varepsilon_{2i} = b + \theta_1 BidLabor_{2i} + \sum_{\theta=2}^{\Theta} \theta x_i + \eta y_{1i} + \varepsilon_{2i}$$
(1)

Here, the $v$'s are systematic utility components, and the $\varepsilon$'s are error terms. $\alpha$ and $b$ are constants to be estimated. $\beta_1$ and $\theta_1$ are parameters associated with bid price, *BidLabor*, and bid working days, *BidLabor*, respectively. $\beta$ and $\theta$ are a vector of parameters related to socio-demographic variables, $x_i$, and $\eta$ is a parameter associated with the observed binary response for WTP for randomly assigned bid prices, $y_{1i}$.

If the two errors in Equation 1 have a joint bivariate normal distribution, we provide the measurement models for the choice indicators in Equation 2 and Equation 3.

$$y_{1i} = \begin{cases} 1 \ if \ U_{1i} > BidCash \\ 0 \ if \ U_{1i} \leq BidCash \end{cases}$$
(2)

$$y_{2i} = \begin{cases} 1 & \text{if } U_{2i} > BidLabor \\ 0 & \text{if } U_{2i} \leq BidLabor \end{cases}$$

(3)

where, $y_{1i}$ and $y_{2i}$ are binary responses for WTP and WTC for randomly assigned bid price and working days, respectively. Here, the joint probability of observing $y_{1i}$ and $y_{2i}$ i.e. $P_{11}$=[ $y_1$=1, $y_2$=1], $P_{10}$=[ $y_1$=1, $y_2$=0], $P_{01}$=[ $y_1$=0, $y_2$=1] and $P_{00}$=[ $y_1$=0, $y_2$=0] is the product of probability of $y_{1i}$ and probability of $y_{2i}$ conditional on probability of $y_{1i}$. Here, $y_{1i}$ is a potentially endogenous variable:

$$L = \prod_{i=1}^{I} \left[ \int_{z_1}^{\infty} \int_{z_2}^{\infty} \varphi(z_1, z_2; \rho) dz_1 dz_2 \right]^{y_1 y_2} \left[ \int_{z_1}^{\infty} \int_{-\infty}^{z_2} \varphi(z_1, z_2; \rho) dz_1 dz_2 \right]^{y_1(1-y_2)} \left[ \int_{-\infty}^{z_1} \int_{z_2}^{\infty} \varphi(z_1, z_2; \rho) dz_1 dz_2 \right]^{(1-y_1)y_2} \left[ \int_{-\infty}^{z_1} \int_{-\infty}^{z_2} \varphi(z_1, z_2; \rho) dz_1 dz_2 \right]^{(1-y_1)(1-y_2)}$$

(4)

where $z_1 = -v_{1i}/\sigma_1$, $z_2 = -v_{2i}/\sigma_2$, $\sigma_1$ and $\sigma_2$ are standard errors, $\phi(z_1, z_2; \rho)$ is the standard bivariate normal distribution, which can be specified as (Poe et al., 1997):

$$\phi(z_1, z_2; \rho) = \frac{\exp-(z_1^2 + z_2^2 - 2\rho z_1 z_2)/2(1-\rho^2)}{2\pi(1-\rho^2)}$$

(5)

where $\rho$ is correlation parameter.

Equation 4 can be estimated using maximum likelihood procedures[8]. We will test for the null hypothesis $\rho = 0$, i.e., the two equations in Equation 1 are independent[9]. If we cannot reject the null hypothesis, the two equations can be estimated using independent univariate probit models and $\phi(.)$ in Equation 5 takes the usual standard univariate normal distribution (Train, 2009).

---

[8] The methodological approach can be generalized for more than two payment vehicles. However, in the case of more than two payment vehicles, we need to find instrumental variables for each payment vehicle to avoid potential endogeneity problems. For example, if we have three payment vehicles, assuming that the order of the questions is the same for all respondents, we can use the traditional hybrid choice model framework (Walker and Ben-Akiva, 2002) with a flexible error structure to solve the problem.

[9] It is a common practice to test of exogeneity assumption between $y_{1i}$ and $y_{2i}$ without incorporating the endogenous ($y_{1i}$) variable as a regressor in the bivariate probit model. However, Filippini et al. (2018) show that the practice may lead to an erroneous conclusion about the correlation of error terms considering the data generation process.

Besides observing the significance of $\rho$, evaluating the coefficient of the potential endogenous variable, $y_{1i}$, between the independent probit models and the bivariate probit model provide an exogeneity test. A substantial change in the magnitude and sign of the parameters of $y_{1i}$ show that the model is not independent and it also confirms that the variable is endogenous (Waters, 1999).

Once addressing the potential endogeneity and correlation issues between the sequences of choices, the ultimate objective of any environmental valuation study is to calculate welfare change. Following Hanemann (1984, 1989), the compensating surplus welfare measure of change is calculated using the following formula:

$$CV^M = \left(V_{1i}^1 - V_{1i}^0\right)/\beta_1$$
$$CV^L = \left(V_{2i}^1 - V_{2i}^0\right)/\theta_1$$
(6)

where $CV^M$ and $CV^L$ are the compensating surplus welfare measure in money and labor. $V_{1i}^1$ and $V_{2i}^1$ are the systematic components of utility if the farmer is WTP and WTC in money and labor, respectively. $V_{1i}^0$ and $V_{1i}^0$ are the status-quo alternatives (doing nothing).

Our survey asked respondents to indicate a representative value that might be paid for labor during the slack and peak agricultural seasons, respectively. However, this value should not be used directly to convert $CV^L$ to monetary units because it does not represent actual wage payments to labor given extant labor exchange mechanisms. Thus, to value labor contributions, we begin with this reported value and adjust it based on previous research. In a similar region of the current study, Kassahun et al. (2020b) reported that the estimated shadow wage rate is 38.63 % of the sample respondent average reported wage rate. In the rural part of Uganda, Meginnis et al. (2020) also reported the shadow wage rate in a range of 15 to 55% of the market wage rate. Similarly, other researchers reported a low estimate of the shadow wage rate in developed countries. For example, Eom and Larson (2006), in their environmental quality valuation study, found the household shadow wage rate is about 70 to 80% of the market wage rate in South Korai. Therefore, in this study, the reported seasonal market wage rate will be adjusted to reflect the shadow wage rate in the calculation of the total compensating surplus welfare measure for accessing reliable irrigation services. We use 0.3863 for the conversion of the market wage rate to the shadow wage rate for the slack and peak agricultural season daily labor cost (Kassahun et al., 2020b). The estimate provides an interval for the lower and upper bound for the shadow wage rate (Equation 7).

$$l, u = \left[ 0.3863 w_{slack}, 0.3863 w_{peak} \right] \tag{7}$$

Note that in the valuation scenario (Text Box 1) and valuation questions (Text Box 2), labor and money are complementary to access reliable irrigation services. Furthermore, the contribution of labor is monthly in the valuation questions. Thus, the total annual compensating surplus welfare measure for accessing reliable irrigation services, $CV^{ML}$, is calculated using Equation 8:

$$CV^{ML} = CV^M + 12 CV^L w \tag{8}$$

where $w$ is an average of the lower and the upper bound of the shadow price provided by Equation 7.

## 5. Results and discussion

### 5.1. Descriptive results and discussion

In addition to data generated from CV scenario and valuation questions, we collected data on socioeconomic and demographic variables hypothesized to have an impact on WTP/WTC for reliable irrigation service. The average gross household cash income is about 6,473 ETB. It is the sum of all marketed agricultural outputs and income from off-farm activities by the household[10]. The average number of economically active household members (working individuals in the household) is about 3.53. We also collected data on the daily labor cost for the slack (14 ETB) and peak (18 ETB) agricultural seasons. Accordingly, from Equation 8, the daily shadow wage rates for the slack and peak agricultural seasons are 5.4 ETB and 6.95 ETB, respectively (Table 2).

To understand the rural labor market, we have asked two important follow-up questions about the shortage and experience of hiring labor (Table 2). About 44% of respondents reported they currently have a labor shortage. However, none of the respondents indicated they had employed wage labor daily for farming activities. Whenever there is a shortage of labor, households usually help each other.

The result of the average per capita income and active household labor (economically active household member per economically dependent household) differences are presented for the three groups of respondents based on their responses for the sequence of the two dichotomous choice CV questions (Figure 1). About 41% of the respondents are willing to contribute to reliable irrigation service in both money and labor (Yes-Yes), 22% only by monetary contribution (Yes-No), and 20% only by labor contribution (No-Yes). As expected, both per capita income and active household labor are highest for the groups that are willing to contribute through both payment vehicles. A comparison of respondents that are willing to pay only in monetary contribution (Yes-No) with the respondent group willing to pay in both payment vehicles (Yes-Yes) shows significantly lower values in the household with active labor (P= 0.0004) but not in the value of per capita income (P= 0.8921). On the other hand, a comparison of the groups willing to contribute only labor (No-Yes) with the

---

[10] The value of marketed agricultural output uses annual average prices for 2007/8 (2000 in the Ethiopian calendar).

**Table 2:** Descriptive Statistics for N = 194: Households Surveyed in the Koga Watershed, and the Expected impact of the variables on willingness to contribute labor (WTC) and willingness to pay (WTP)

| Variable | Description | Mean | Std. Dev. | Expected sign |
|---|---|---|---|---|
| $y_{2i}$ | 1 if household is willing to contribute the proposed bid working days for reliable irrigation service, otherwise 0 | 58% | | |
| *BidLabor* | Bid number of working days per month | 1.99 | 0.71 | - |
| Land per household | Land holding per household size (*kada*/ household size) | 1.04 | 0.54 | - |
| Experience with irrigation | 1 if household head indicated s/he has practical irrigation farming experience, otherwise 0 | 18% | | + |
| Working household members | Number of economically active individuals in a household i.e., working-age population age above 14 and below 66 years of age | 3.53 | 1.64 | |
| Dependency ratio | Ratio of dependents (people younger than 15 or older than 65) to the working-age population | 0.86 | 0.63 | - |
| Farm cart ownership | 1 if a household owns a farm cart either with horse or mule or both, and 0 otherwise | 31% | | + |
| Young household head | 1 if age of household head is less than 43, otherwise 0 (division based on mean age) | 55% | | + |
| Education | Highest formal schooling completed by any household member in years | 5.65 | 4.15 | + |
| Gross household income | Sum of all marketed agricultural outputs plus income from off-farm activities by all members of the household | 6473 | 5943 | |
| Per capita income | Gross household income divided by household size | 1123 | 1462 | + |
| $y_{1i}$ | 1 if household is willing to pay the proposed bid price (cash) for reliable irrigation service, otherwise 0 | 63% | | - |
| *BidCash* | Bid price | 44.44 | 14.46 | |
| Labor cost, slack period | Average slack agricultural season daily labor cost in ETB | 13.55 | 2.53 | |
| Labor cost, peak period | Average peak agricultural season daily labor cost in ETB | 17.71 | 2.62 | |
| Shadow wage, slack | Estimated average slack agricultural season daily shadow wage rate in ETB | 5.23 | 0.97 | |
| Shadow wage, peak | Estimated average peak agricultural season daily shadow wage rate in ETB | 6.84 | 1.01 | |
| Labor shortage | Do you currently have labor shortage for crop and livestock farming | 44% | | |
| Hiring labor | Have you ever employed wage labor daily for farming activities | 0% | | |

NOTE: ETB indicates Ethiopian Birr. At the time of the survey, 1 ETB was equal about 0.1 USD.

groups willing to pay only in money (Yes-No) shows a significantly lower value of per capita income (P= 0.0012). However, there is no significant difference in the amount of labor between the two groups (P=0.4171). The implication is that absolute and relative endowment of labor and income rather than strategic or protest response profoundly influenced respondents' choices. That means there is no behavioral response anomaly regarding the standard assumptions. More importantly, for this study, besides the significant proportion of individuals willing to contribute in both payment vehicles, the incorporation of labor as a payment vehicle increases flexibility for the 20% low-income households in our sample to more accurately express their preferences about reliable irrigation service. This is a valid conclusion considering the significant and positive effect of income on willingness to pay (Table 3).

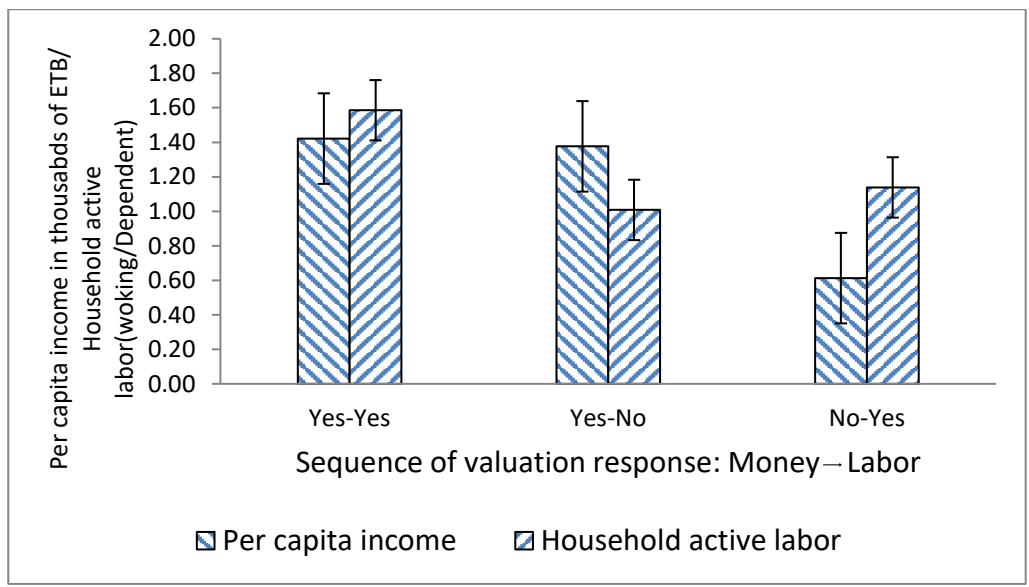

**Figure 1:** Respondent relative endowment of income and labor and their responses from the sequence of valuation question in the order of WTP and WTC

Another significant behavioral response anomaly that needs attention in valuation study is related to the issue of anchoring or starting point bias. For appropriate model specification, we have made expletory descriptive analysis with the responses of valuation questions. We test significant differences between means of maximum WTP/WTC from the follow-up question grouped by starting bids from the dichotomous CV questions (Figure 2). At first both the average maximum WTP and WTC seems increasing along with the starting bids. Nevertheless, there is no statistical difference in the average maximum WTP and WTC with the 95% confidence intervals with starting bids.

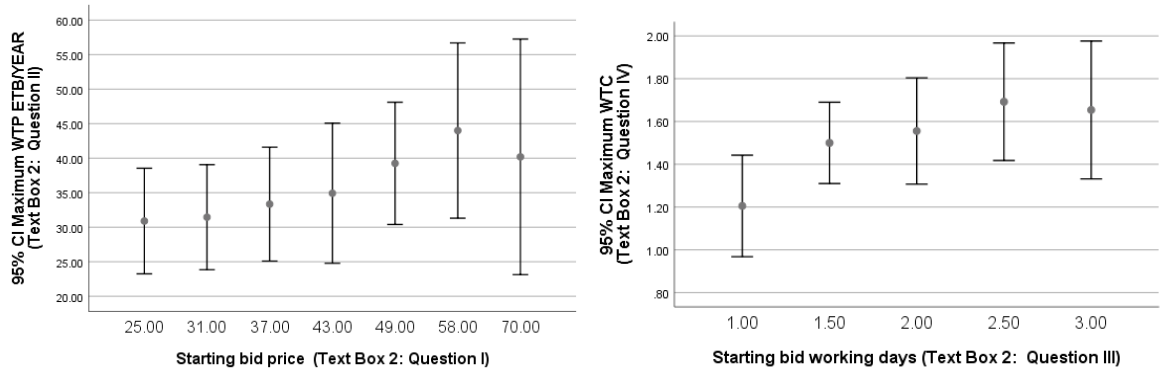

**Figure 2:** Significant differences test between average maximum WTP (left) and WTC (right) grouped by starting bids from the dichotomous CV questions.

For testing cross payment vehicle starting point bias, we categorized the maximum willingness to work values by bid price (Figure 3). The results show that there are no significant differences in the mean maximum WTP/WTC with starting bid price. That means there is no order effect/ starting point bias.

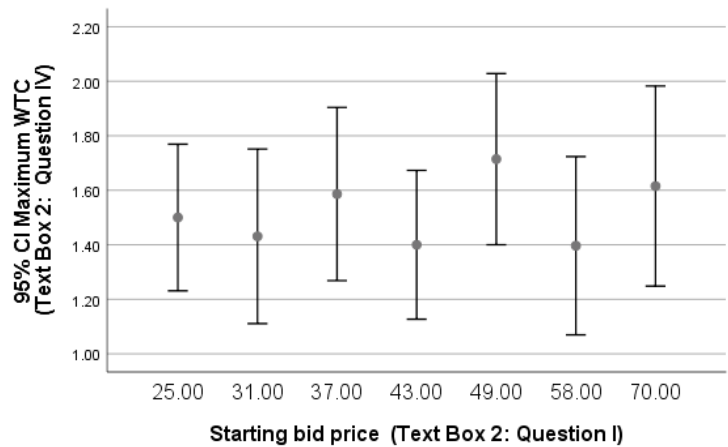

**Figure 3:** Significant differences test between average maximum WTC grouped by starting bid price.

**Table 3:** Estimated Coefficients for Univariate and Bivariate Probit Models

| Equation, Variable | Univariate Probit | | Bivariate Probit | |
|---|---|---|---|---|
| | Coefficients | t | Coefficients | t |
| *Dependent $y_2$* | | | | |
| $y_1$ | 0.09 | 0.36 | -1.21 | -7.66 |
| BidLabor | -0.98 | -5.45 | -0.73 | -5.46 |
| Land per household | -0.28 | -1.27 | -0.18 | -1.02 |
| Experience with irrigation | 0.34 | 1.21 | 0.71 | 2.48 |
| Dependency ratio | -1.02 | -4.71 | -1.06 | -5.84 |
| Farm cart ownership | 0.41 | 1.72 | 0.32 | 2.00 |
| Young household head | 1.25 | 4.75 | 1.49 | 6.68 |
| Education | 0.02 | 0.83 | 0.07 | 2.56 |
| Constant | 2.37 | 4.64 | 2.19 | 4.93 |
| *Dependent $y_1$* | | | | |
| BidCash | | | -0.04 | -5.71 |
| Dependency ratio | | | -0.89 | -4.09 |
| per capita income | | | 0.55 | 4.14 |
| Experience with irrigation | | | 1.17 | 3.14 |
| Young household head | | | 0.84 | 3.67 |
| Education | | | 0.07 | 2.48 |
| Constant | | | 1.43 | 3.18 |
| Athrho* | | | 12.95 | 11.4 |
| rho** | | | 1 | |
| Log likelihood | -85.67 | | -162.31 | |
| Chi-squared | 68.04 | | 147.61 | |
| P | 0.00 | | 0.00 | |
| N | 194 | | 194 | |

\* is the Fisher's Z transformation of the correlation.
\*\*Likelihood-ratio test of rho=0: chi2(1) = 130.027 Prob > chi2 = 0.0000

## 5.2. Econometric results and discussion

Estimates of the univariate probit model for labor contribution and the bivariate probit model for the joint cash and labor contributions are in Table 3. The likelihood-ratio test for $\rho = 0$ for the joint model is significant (P=0.0001), so we can reject the null hypothesis that the decisions about money and labor contribution for reliable irrigation service are exogenous. Consequently, the two models in Equation 1 should not be estimated using two separate univariate probit models. Furthermore, we run an additional test of exogeneity by observing the sign and the significant levels of the response of WTP decision, $y_{1i}$, on WTC for reliable irrigation service decision, $y_{2i}$, between the two probit models (Table 3). The result shows that the response of WTP is positive and insignificant for the case of the univariate probit model. Whereas, it is negative and significant for the bivariate probit model. The result further strengthens the hypothesis that the response of the WTP decision is endogenous for WTC. Thus, the bivariate probit model would be a better option for accounting both the correlation of the sequence of choices and the endogeneity problem. For the rest of our discussion, we focus on the results of the bivariate probit model.

One of the fundamental reasons for the incorporation of labor time contribution as a payment vehicle for the valuation of environmental services in developing countries is that scarcity of cash exchanges may lead to underestimation of the value of ecosystem services. Therefore, it is hypothesized that including labor contributions as a payment vehicle gives more flexibility for the majority of the poor in rural areas to reveal their preferences (Schiappacasse et al., 2013). Our result confirms this hypothesis. Holding the effect of other variables constant, the probability of an individual's willingness to contribute labor (WTC) for reliable irrigation service will increase by 37.8 % if the respondent chooses not to pay ($y_{1i} = 0$) a monetary contribution (Table 4)[11]. However, this result alone does not provide evidence that including labor as a payment vehicle increases flexibility for the low-income residents of rural areas to reveal their preference. Thus, we have done further analyses to identify whether the relationship is due to the strategic response or consistent with the hypothesis of greater flexibility (Figure 1, Section 5.1). We affirm that the incorporation of labor indeed increases flexibility for low-income households in our sample.

---

[11] For computing the marginal effect, both significant and insignificant variables are considered. However, in Table 3, only the marginal effects of significant variables presented.

**Table 4:** Average Marginal effects (AME) of explanatory variables on WTC ( $y_{2i}$ ), N=194

| Variable | AME | Std. Err. | 95% C.I. | | Mean for AME computation |
|---|---|---|---|---|---|
| $y_{1i}$ * | -0.38 | 0.06 | -0.50 | -0.26 | 0.00 |
| BidLabor | -0.16 | 0.04 | -0.24 | -0.09 | 1.99 |
| Experience with irrigation | 0.20 | 0.10 | 0.01 | 0.39 | 0.18 |
| Dependency ratio | -0.24 | 0.05 | -0.34 | -0.13 | 0.86 |
| Farm cart ownership* | 0.08 | 0.05 | 0.00 | 0.17 | 0.00 |
| Young household head* | 0.52 | 0.07 | 0.39 | 0.65 | 0.00 |
| Education | 0.02 | 0.01 | 0.00 | 0.03 | 5.65 |

(*) AME is for discrete change of dummy variable from 0 to 1

As the standard theory of demand suggests, both the bid price ( *BidCash*) and the bid working days ( *BidLabor*) are highly significant determinants of WTP and WTC for reliable irrigation service decisions. The higher the bid price and the number of working days, the lower the probability of an individual WTP and WTC for reliability irrigation service (Table 3).

Other than the bid working day, the dependency ratio is a significant but negative determinant of WTP and WTC values (Table 3). A higher dependency ratio may decrease the probability of WTC for reliable irrigation service because caring for dependents and accomplishing basic household tasks take considerable labor time (Shiferaw and Holden, 1998). Similarly, a higher dependency ratio means a lower available disposable income for WTP.

Agriculture is still considered the most labor-intensive sector in developing countries. Moreover, a lack of road infrastructure and means of transportation in rural areas of developing countries can increase the amount of time required to accomplish various agricultural activities. Having a farm cart with pack animals reduces the amount of time required to transport various agricultural materials and inputs considerably. Hence, it increases the probability of WTC for reliable irrigation services (Table 3).

The other variable that has a highly significant and positive impact on the WTP and WTC for reliable irrigation service is being a young household head (Table 3). The magnitude of the parameter for being a young household head is higher in the case of labor contribution (1.49) compared to cash (0.84) contribution. The marginal effect of being young increases the probability of WTC for reliable irrigation service by 52% (Table 4). The result is consistent with the findings of labor-intensive projects in developing countries, where younger people likely to contribute to labor (Kassahun and Jacobsen, 2015; Kassahun et al., 2020b). Similarly, education has a positive effect on the WTP and WTC for reliable irrigation services. Nevertheless, an increase in the year of education leads to a small marginal increment on the WTP and WTC for reliable irrigation service compared to other variables of interest.

Another most crucial variable for the decision for WTP is irrigation farming experience. It is a highly significant determinant of both money and labor contribution (Table 3). There is a substantial difference in the average maximum WTP between respondents with irrigation experience and those without experience. Individuals with irrigation farming experience are willing to pay 1.6 times of farmers that do not have irrigation farming experience (see table below). The reason is more likely due to the high expectation of the productivity and profitability of irrigation farming realized from irrigation farming experience (Kassahun et al., 2016). However, there is no statistically significant difference between the groups with experience or not for irrigation for labor contributions (Figure 4).

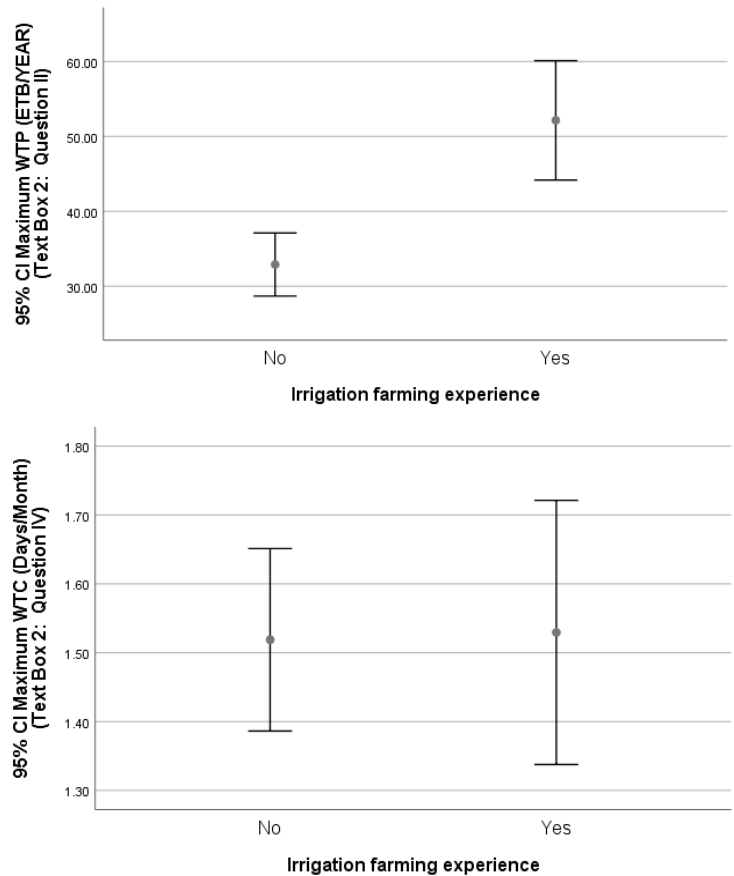

**Figure 4:** Average maximum WTP (top) and WTC (bottom) between respondents with irrigation farmers experience

## 5.3. WTP and WTC for access to reliable irrigation service

Thus far, we have examined the sign and magnitude of the determinants of WTP and WTC for relevant explanatory variables. However, the total value of reliable irrigation services is also essential. Our study differentiates the purpose of money and labor contribution for managing reliable irrigation service in the framework of watershed management (Text Box 1 and Text Box 2). The contributions in both types of payment vehicles are complementary for access to reliable irrigation water. However, if the respondents consider upstream and downstream management activities as different bundles of services, some caveats may apply to the relative size of the money/labor contribution. In this case, the respondents may reveal a different WTP/WTC regardless of the amount of cash and labor available to them. However, we do not consider this as an issue in our case. That is because our result shows that both absolute and relative endowment of labor and income rather than strategic response influenced respondents' choices (Figure 1).

The mean WTP for irrigation service is about 57 ETB (about 5.7 USD at the time of the survey) per Kada of irrigable land per year and 29 working days of labor per year per Kada of irrigable land (Table 5). The average monetary equivalent of labor contribution is equal to 177.82 ETB per Kada of irrigable land per year with seasonal daily shadow wage rate range of 5.23 to 6.84 ETB for slack and peak agricultural seasons. Hence, the average annual willingness pay for access to reliable irrigation service is 235 ETB (23.5 USD) per Kada of irrigable land (Equation 8). Of the total average annual willingness pay for access to reliable irrigation service, cash contribution comprises between 23.57 to 27.6 %. Here, the total value of reliable irrigation service could be underestimated if reliable irrigation service would have been estimated using only monetary contribution.

**Table 5:** Mean WTP and WTC for reliable irrigation service (N=194)

| Valuation Measure | Mean willingness to contribute | Standard dev. |
|---|---|---|
| WTP (ETB /Year) | 57.37 | 31.68 |
| WTC (Days/year) | 28.77 | 14.85 |
| Slack agricultural season WTC (ETB /Year) | 150.46 | 77.65 |
| Peak agricultural season WTC (ETB /Year) | 196.76 | 101.55 |
| Average WTC (ETB /Year) | 177.82 | 91.76 |
| Total average annual willingness to contribute (ETB /Year) | 235.19 | 110.58 |

Note: At the time of the survey, 1 ETB was equal about 0.1 USD.

Furthermore, using our estimate of 57 ETB/ Kada of irrigable land per year, the aggregate monetary contribution for the total irrigation command area of 7000 ha is 1,596,000.00 ETB/year. The estimate is about 90 % of the budget allocated over six years (2002-2007) for soil conservation and reforestation to reduce sediment by 50 % during the construction of the Koga reservoir[12]. Therefore, in absolute terms, the monetary contribution is not considered low for this study.

## 6. Conclusions and recommendations

Considering the limited importance of the cash economy in rural areas, many researchers in developing countries propose WTC as a means of welfare measurement that complements the WTP value. This paper contributes with empirical evidence for the hypothesis that including money and labor in valuation studies improves estimates of the demand for environmental goods for respondents with different endowments of income and labor. We demonstrate this through a rural household contingent valuation survey designed to elicit the value of access to reliable irrigation water in Ethiopia. Furthermore, the methodological approach employed can be generalized for more than two payment vehicles. However, in the case of more than two payment vehicles, we need to find instrumental variables for each payment vehicle to avoid potential endogeneity problems. Thus, considering ours and other previous studies of multiple payment vehicles in developing country contexts, here we outline four conditions for how, when, and why multiple payment vehicles would be appropriate for valuation studies.

First, if there is a clear indication of variation in the use of different payment vehicles across different geographical localities, a spatial split-sample design with payment vehicles is necessary. The number of payment vehicles depends on the disaggregation geographical localities. However, in a situation with varying levels of familiarity with the payment vehicles, a follow-up question might be needed to investigate bias related to payment vehicles post data collection. Under this condition, any comparison of valuation studies with a split-sample design between different payment vehicles must rely on a shadow price as market price information may not be available.

---

[12] https://www.afdb.org/fileadmin/uploads/afdb/Documents/Project-and-Operations/Ethiopia_-_Koga_Irrigation_and_Watershed_Management_Project_-_Appraisal_Report.pdf

Second, variation in institutional trust/distrust may be another factor for justifying adding multiple payment vehicles in a local area. For example, if a (potentially corrupt) local municipality proposes the construction of a recreational facility with a fund from a local community, individuals' willingness to pay could be low because of institutional distrust. In this situation, researchers should select payment vehicles that are less prone to corruption. However, this may be challenging if alternative payment vehicles are not obvious. The use of focus group discussion or pre-test, as we did for the monetary contribution for the current study, can be a way to find a payment vehicle that is trustworthy and consequential.

Third, when multiple payment vehicles are introduced in a sequence of valuation questions, unintended behavioral response anomalies (e.g., strategic response, protest response, staring point bias, or anchoring effect) might occur that could affect the valuation result. Focus group discussions and pre-tests should be carefully carried out to minimize these effects before administrating a final survey. Furthermore, a post-data-collection test should be conducted to select an appropriate modeling framework for accounting these behavioral response anomalies if there is an issue. Alternatively, a CE format may be used, in which multiple payment vehicles can be estimated simultaneously.

Fourth, multiple payment vehicle introductions in a sequence of valuation questions might also lead to correlation and endogeneity issues. This article provides empirical evidence that the assumption of exogeneity in the sequence of WTP and WTC valuation questions leads to biased estimated coefficient and erroneous conclusions regarding the effect of cross-payment vehicle trade-offs. The implication is also valid for a choice experiment (CE) survey that allows the respondent to choose payment vehicles and administer the CE survey using the selected payment vehicle.

Finally, given the comprehensive tests made in our study, we conclude that limited cash availability and lack of future income realization through irrigation farming are the main plausible explanation for low WTP in monetary payment. Of the 235 ETB total average annual willingness pay for access to reliable irrigation service, cash contribution comprises between 23.57 to 27.6 %. The implication is that socially desirable projects might be rejected based on cost-benefit analysis with a welfare gain underestimation due to mismatch of payment vehicles choice in valuation studies. Thus, including payment vehicles that are currently available to the farmer and less prone to corruption are essential for the assessment of demand in developing countries.

**Acknowledgements**

We thank the Cornell University Integrated Watershed Management and Hydrology Program for funding this research through a gift of Cornell donor providing funds for research in Africa. ***Additional acknowledgement will be included later**